\documentclass[11pt,letterpaper]{article}
\usepackage[margin=1in]{geometry}
\usepackage{microtype,caption,titlesec,placeins,needspace}
\titleformat{\section}{\large\bfseries}{\thesection}{.7em}{}
\titleformat{\subsection}{\normalsize\bfseries}{\thesubsection}{.7em}{}
\usepackage[T1]{fontenc}
\usepackage[utf8]{inputenc}
\usepackage{amsmath,amssymb,amsthm,graphicx,booktabs,array,url,algpseudocode,float}
\usepackage[colorlinks=true,linkcolor=black,citecolor=blue!50!black,urlcolor=blue!50!black]{hyperref}
\usepackage{xcolor}
\newcommand{\W}{\mathcal{W}}
\newcommand{\SP}{S_{\pi}}

\newcounter{paperalgorithm}
\newfloat{algorithmfloat}{tbp}{loa}
\newenvironment{paperalgorithm}[2]{%
\begin{algorithmfloat}[!htbp]\small%
\refstepcounter{paperalgorithm}\label{#2}%
\hrule\smallskip\textbf{Algorithm \thepaperalgorithm. #1}\par\smallskip\small
}{\smallskip\hrule\end{algorithmfloat}}
\algrenewcommand\algorithmicrequire{\textbf{Input:}}
\algrenewcommand\algorithmicensure{\textbf{Output:}}
\graphicspath{{figures/}}
\newcommand{\paperkeywords}{Phase unwrapping, adaptive tiling, quadtree, kd-tree, domain decomposition, least-squares reconstruction, Poisson equation, discrete cosine transform, image processing, remote sensing, runtime analysis, computational complexity}
\hypersetup{pdftitle={Adaptive Tiling for Least-Squares Phase Unwrapping: Runtime and Accuracy},pdfauthor={Antoine Moevus and Max Mignotte},pdfkeywords={\paperkeywords}}
\begin{document}
\raggedbottom
\widowpenalty=10000
\clubpenalty=10000
\begin{center}
{\Large Adaptive Tiling for Least-Squares Phase Unwrapping\par}
\vspace{.4em}
{\large Runtime and Accuracy\par}
\vspace{1.2em}
Antoine Moevus \quad Max Mignotte\\[.4em]
Department of Computer Science and Operations Research\\
Universit\'e de Montr\'eal\\[.7em]
\textbf{Technical Report}\\[.7em]
September 2026
\end{center}
\vspace{.7em}
\begin{abstract}
Phase unwrapping estimates the missing multiples of $2\pi$ in measured phase images. For large images, tiling limits the size of local reconstruction problems and enables parallel processing. Adaptive tiling could further reduce the number of local problems and boundaries by retaining large tiles where little refinement is needed. We investigate whether this reduction makes reconstruction faster. We compare complete reconstruction time and accuracy for a regular grid, quadtree, and kd-tree partitions. We also evaluate nine criteria for deciding where quadtree tiles should be subdivided, including residue count, fringe density, and measures of phase variation, at different tile sizes and budgets. In single-threaded experiments on a heterogeneous image dataset, optimized adaptive partitions use fewer tiles but remain slower than the optimized grid, and some reconstructions lose substantial accuracy. Stage measurements explain why: constructing the partition and solving larger retained tiles outweigh the savings at tile boundaries. The criterion comparison also shows that more refinement does not consistently improve accuracy. These results motivate evaluating adaptive partitions by the complete time needed to reach a chosen reconstruction accuracy, including whether limited refinement can provide a faster approximate result.
\end{abstract}
\noindent\textbf{Keywords:} \paperkeywords.

\section{Motivation and scope}
Phase unwrapping estimates the missing multiples of $2\pi$ in a measured angle image. For large images, tiling reduces the size of individual reconstruction problems. Tiles can be processed sequentially to limit memory use, or in parallel before their solutions are assembled \cite{chen2002phase}. The choice of partition must therefore account for both local reconstruction and boundary reconciliation.

Several phase-unwrapping methods already separate local reconstruction from the assembly of neighboring regions. Strand et al.\ \cite{strand1999two} unwrap small square blocks under the assumption that the true phase range within each block is below $2\pi$, then align them by least-squares fitting of integer-cycle offsets. Antonopoulos et al.\ \cite{antonopoulos2015tilebased} combine polynomial fits within tiles with reliability-guided merging for digital holography. The tiled extension of SNAPHU unwraps tiles independently, subdivides them into reliable regions, and estimates region offsets through a second network optimization \cite{chen2002phase}.

Here we study tiled least-squares reconstruction. Least-squares methods fit a phase field to differences between neighboring observations \cite{ghiglia1994robust,ghiglia1998two}; Moevus and Mignotte \cite{moevus2026tilebased} use local Poisson solves in a fixed-grid formulation. A regular grid uses a fixed tile size, whereas an adaptive partition retains larger tiles where refinement appears unnecessary. Adaptive regions also have precedents: Baldi \cite{baldi2001twodimensional} subdivides using phase jumps and joins regions through interface-error minimization, while Yu et al.\ \cite{yu2011residues} construct regions from residue clusters.

\begin{figure}[!ht]
\centering\includegraphics[width=.94\textwidth]{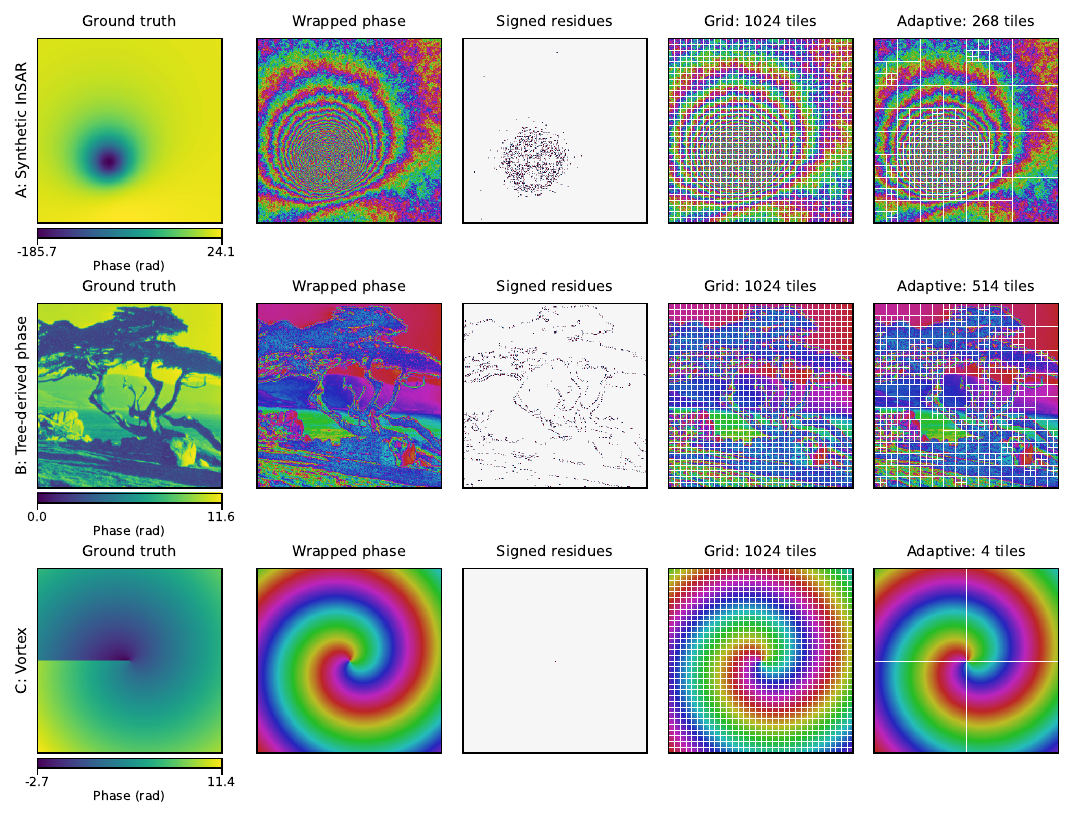}
\caption{Regular and adaptive partitions on three $256\times256$ phase fields: A, a generated interferometric synthetic aperture radar (InSAR) field; B, a field derived from tree-photograph intensities; C, a generated vortex field. Columns show the known generating phase (ground truth, GT), wrapped phase, signed cell residues (red: $+1$, blue: $-1$, white: zero), a grid of $8\times8$-pixel tiles, and a residue-driven quadtree. Each GT panel has its own color scale in radians; wrapped panels share the cyclic scale $[-\pi,\pi]$. White lines mark tile boundaries. The quadtree has minimum side $s=8$ pixels and stopping budget $B=512$ leaves. It retains 268, 514, and four tiles in A--C, versus 1024 grid tiles in each case. A four-way split adds three leaves, allowing the budget to be exceeded by two. These unquantized examples illustrate allocation, not reconstruction accuracy or runtime.}\label{M-fig:overview}
\end{figure}

Figure~\ref{M-fig:overview} illustrates the attraction of adaptivity: many small tiles can be replaced by a few large regions. Whether this saves time depends on the cost of identifying and reconstructing those regions, as well as the boundaries removed.

Our question is whether adaptive allocation improves complete reconstruction time or reference agreement when the local solver and joining rule are fixed. We first compare a regular grid with four-way and binary partitions using residue-driven refinement. We then examine how alternative criteria and restricted tile budgets affect allocation and error. This separates the runtime value of the tested adaptive pipeline from the broader question of where refinement should be concentrated.

\section{Reconstruction with a common solver}\label{M-sec:method}
\begin{figure}[!htbp]
\centering\includegraphics[width=\textwidth]{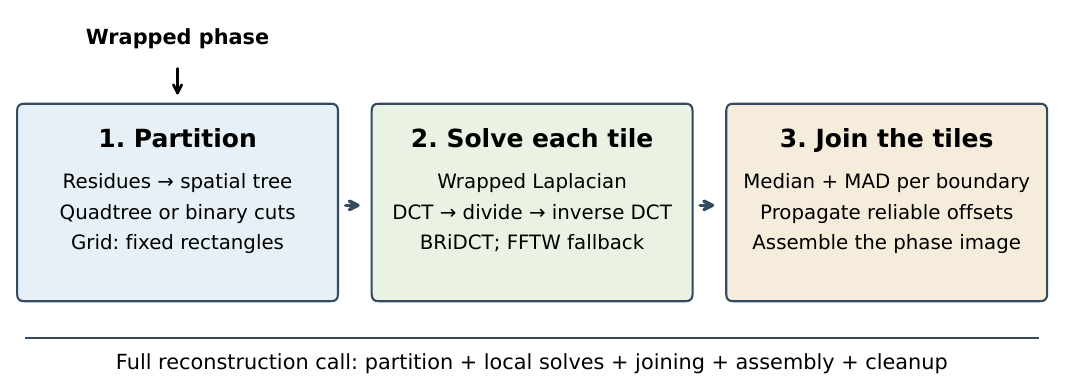}
\caption{Common pipeline for grid, quadtree, and kd-tree reconstruction. The partition sets the tile boundaries; local Poisson solves use the BRiDCT discrete cosine transform (DCT), with FFTW for unsupported shapes. Seam discrepancies are summarized by their median and median absolute deviation (MAD). A separate reliability tree connects the reconstructed tiles and determines their relative phase offsets. The complete-call timer includes preparation, all reconstruction stages, and temporary-buffer cleanup. Transform plans are reused in the main timing comparison; first-call latency is evaluated separately.}\label{M-fig:pipeline}
\end{figure}

Figure~\ref{M-fig:pipeline} separates partition construction, local reconstruction, and boundary reconciliation. A \emph{spatial tree} determines tile boundaries. A different tree, introduced in Sec.~\ref{M-sec:joining}, connects the reconstructed tiles to propagate their offsets. Keeping these roles separate allows the grid, quadtree, and kd-tree to use the same local solver and joining rule.

\subsection{Residues and subdivision}\label{M-sec:partition}
Let $\psi$ denote the wrapped observation and $\W$ reduce an angle to $(-\pi,\pi]$. A cell consists of four neighboring pixels arranged in a square. Compute the principal differences $\W(\psi_q-\psi_p)$ along its four edges and sum them with signs following a consistent orientation around the cell. A nonzero sum, measured in multiples of $2\pi$, is a \emph{residue}: the four observed differences cannot all belong to one consistent phase field. For the main runtime comparison, the tile score is the sum of absolute residue charges over cells wholly inside it, excluding cells crossed by its boundary. Alternative scores are examined in Sec.~\ref{M-sec:criteria}.

\begin{figure}[!ht]
\centering\includegraphics[width=.92\textwidth]{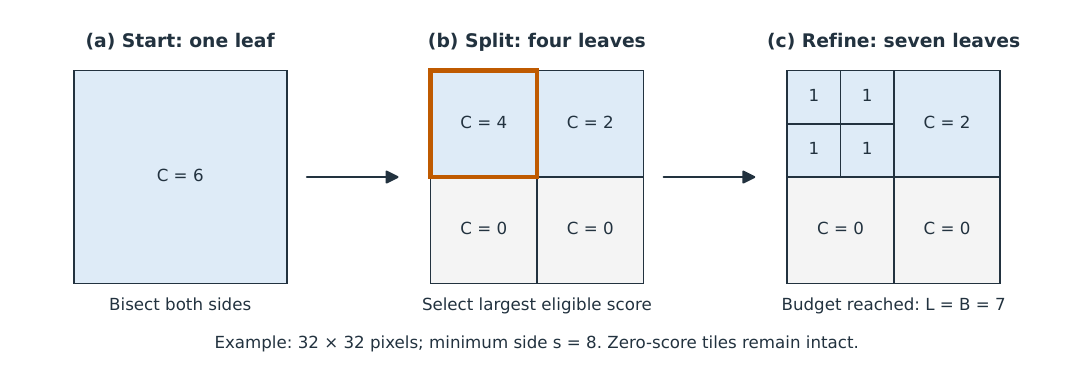}
\caption{Priority-based quadtree refinement on a $32\times32$ image, with minimum tile side $s=8$ pixels and stopping budget $B=7$ leaves. $C$ is an illustrative nonnegative tile score, and $L$ is the leaf count. (a) The root covers the image. (b) Bisecting both sides creates four $16\times16$ children; the orange outline selects the highest-scoring eligible child. (c) Splitting that child creates four $8\times8$ tiles and brings the total to seven leaves. Refinement stops at the budget even though another tile remains eligible with positive score. Zero-score tiles are retained.}\label{M-fig:quadtree-schematic}
\end{figure}

The quadtree starts with the image rectangle. A split bisects both sides and replaces one terminal tile, or \emph{leaf}, with four children (Fig.~\ref{M-fig:quadtree-schematic}). Both sides must be at least $2s$, where $s$ is the minimum permitted tile side. With a tile budget, the largest-scoring eligible leaf is split first. Refinement stops when the budget is reached or no eligible positive-score leaf remains. The final runtime comparison removes the budget restriction: every eligible tile containing a residue is split. In that setting, depth-first traversal produces the same leaves without maintaining their priority in a queue.

\begin{samepage}
With a consistent oriented-edge convention, a residue-free rectangle admits an exact fit to its observed differences in exact arithmetic: zero circulation around every cell makes the integral between two pixels independent of the path. This is a local consistency statement, not a guarantee of the true phase. For example, slopes of $3\pi/2$ and $-\pi/2$ per pixel give the same wrapped observations and no residues. Recovering the physical phase requires additional sampling assumptions \cite{itoh1982analysis}. Inconsistencies can also lie across tile boundaries and therefore escape each tile's internal test.
\par\end{samepage}

\subsection{Binary alternatives to the quadtree}\label{M-sec:geometry}
A kd-tree splits a rectangle into two children with one horizontal or vertical cut. It can refine one direction without refining the other, which may suit elongated structures. The cut is eligible only if both children retain at least $s$ pixels along the divided axis. We test \emph{KD midpoint}, which bisects the longest eligible side, and \emph{KD dyadic}, which places the cut at the largest power of two no greater than half that side, clamped to respect the minimum size. The latter favors dimensions supported by fast transforms without guaranteeing that every resulting tile has power-of-two dimensions.

Both kd-trees use the same residue stopping test as the quadtree, so their comparison measures the effect of cut geometry and the resulting tile shapes. We also examine a residue-balanced rule, which places the cut near the median of the residue distribution instead of dividing side lengths. Its measured cost and accuracy are reported with the other geometric comparisons in Sec.~\ref{M-sec:results}.

\subsection{Local reconstruction and boundary reconciliation}\label{M-sec:joining}
Let $g_T$ collect the observed principal differences inside tile $T$, and let $D_T$ compute neighboring differences of a candidate phase $u$. The local least-squares problem is
\begin{equation}
 u_T\in\arg\min_u\|D_Tu-g_T\|_2^2.\label{M-eq:local}
\end{equation}
When the observed differences are inconsistent, this fit finds a compromise between them; a small fitting residual does not by itself guarantee agreement with the reference phase. The normal equations form a discrete Poisson system with no connections outside the tile. A discrete cosine transform (DCT) diagonalizes this system: transform the right-hand side, divide by the nonzero eigenvalues, and apply the inverse transform \cite{ghiglia1994robust}. The constant coefficient is set to zero because local differences do not determine an additive offset.

The complete-runtime comparison in Sec.~\ref{M-sec:runtime} uses BRiDCT~\cite{moevus2026bridct}, our high-performance single-precision DCT implementation, for both the grid and adaptive partitions. Based on the Shao--Johnson factorization \cite{shao2008typeii}, it supports square and rectangular tiles with power-of-two side lengths from 8 to 1024 pixels. This range is particularly useful for quadtree reconstruction, where repeated bisection produces tiles of widely varying sizes. Other shapes use cosine transforms from the FFTW library. Transform plans, which store reusable setup information, and working buffers are cached by shape.

After each local solve, phase values are rounded to multiples of $\delta=2\pi/256$ radians, using nearest-value rounding with ties to even. The remaining additive offsets must then be reconciled across shared boundaries. We follow the fixed-grid boundary-difference construction of Moevus and Mignotte \cite{moevus2026tilebased}, and account for unequal seam lengths in the median/MAD reliability rule below.

Let $u_a,u_b$ be the local solutions on neighboring tiles, with unknown additive offsets $o_a,o_b$. A \emph{seam} is their shared boundary. For adjacent pixel pairs $(p_k,q_k)$ crossing it from tile $a$ to tile $b$, form
\begin{equation}
 d_k=u_a(p_k)+\W(\psi(q_k)-\psi(p_k))-u_b(q_k).\label{M-eq:seam}
\end{equation}
Its median $m_{ab}=\operatorname{median}_k d_k$ estimates $o_b-o_a$. The median absolute deviation (MAD), $D_{ab}=\operatorname{median}_k|d_k-m_{ab}|$, measures disagreement along the seam rather than the size of its offset. With seam length $n_{ab}$ (the number of pixel pairs), define
\begin{equation}
 w_{ab}=\frac{n_{ab}}{(1.4826D_{ab}+\eta)^2},
 \label{M-eq:weight}
\end{equation}
where $\eta=\delta$ is one phase-quantization step and keeps the weight finite when the seam discrepancies agree exactly. The factor 1.4826 expresses MAD on a Gaussian standard-deviation scale. The seam-length factor is heuristic: correlated boundary samples do not justify treating this weight as calibrated uncertainty.

Reliability-tree joining treats tiles as graph vertices and shared boundaries as graph edges. A maximum-weight spanning tree connects every tile without cycles while maximizing the sum of selected reliability weights. Median offsets are propagated from a fixed root along this tree. This joining tree is distinct from the spatial subdivision tree: it determines relative phase offsets, not tile boundaries.

\subsection{Partition and reconstruction algorithms}
Algorithms~\ref{M-alg:partition} and~\ref{M-alg:reconstruct} separate the decisions that are varied from the operations held fixed. The score-map builder $A$ selects where refinement is useful, while the split rule $G$ determines its geometry. Section~\ref{M-sec:criteria} defines the candidate scores; Sec.~\ref{M-sec:geometry} defines the binary alternatives.

Algorithm~\ref{M-alg:partition} uses a summed-area table to obtain each rectangular score from four cumulative entries. It splits the highest-scoring eligible tile while retaining zero-score and ineligible leaves. Checking the budget before a split permits an overshoot of two leaves for quadtrees, but none for binary trees. With no budget limit, depth-first traversal gives the same leaves without a priority queue.

\begin{paperalgorithm}{Priority-based adaptive partition}{M-alg:partition}
\begin{algorithmic}[1]
\Require Wrapped phase $\psi$; nonnegative cell-score builder $A$; split rule $G$; minimum side $s$ (pixels); stopping budget $B\geq1$ (integer or $\infty$).
\Ensure Rectangular leaf partition $\mathcal L$ covering the image.
\State Compute $A(\psi)$ and its summed-area table
\State Define $C(T)$ as the sum of scores over cells inside tile $T$
\State Initialize $\mathcal L$ with the image rectangle and queue it if eligible with positive score
\While{$|\mathcal L|<B$ and an eligible positive-score leaf exists}
 \State Select the eligible leaf $T$ with largest $C(T)$
 \State Replace $T$ in $\mathcal L$ with its children under $G$
 \State Compute child scores and update the priority queue
\EndWhile
\State \Return leaves in fixed order
\end{algorithmic}
Eligibility enforces the minimum side. Ties use insertion order; children cover their parent without overlap. Zero-score and ineligible tiles remain leaves. Set $B=\infty$ for unrestricted refinement; a finite budget can be exceeded by at most two leaves for a quadtree, and is not exceeded by binary splits.
\end{paperalgorithm}

\begin{samepage}
Algorithm~\ref{M-alg:reconstruct} accepts either this adaptive partition or a regular grid. It solves the local problems, estimates seam offsets, and connects the tiles in decreasing order of reliability without creating cycles. A fixed root supplies the additive reference; assembly, minimum subtraction, and rounding complete the output. Ground truth is used only for subsequent accuracy evaluation.\par\end{samepage}

\begin{paperalgorithm}{Local reconstruction and tile reconciliation}{M-alg:reconstruct}
\begin{algorithmic}[1]
\Require Wrapped phase $\psi$; rectangular tiling $\mathcal L$; fixed quantization convention.
\Ensure Reconstructed phase $\widehat\phi$ with a common additive reference.
\For{each tile $T\in\mathcal L$}
 \State Solve (\ref{M-eq:local}) by DCT and apply local quantization
\EndFor
\State Find neighboring tiles and their shared boundary samples
\State Compute seam medians $m_{ab}$ and weights $w_{ab}$ by (\ref{M-eq:seam})--(\ref{M-eq:weight})
\State Select a maximum-weight spanning tree of neighboring tiles
\State Fix one root offset to zero; propagate $o_b-o_a=m_{ab}$ along the tree
\State Assemble $\widehat\phi|_T=u_T+o_T$
\State Shift the output minimum to zero, quantize, and return $\widehat\phi$
\end{algorithmic}
Tiles cover the image without overlap. The single-tile case has offset zero and no seams. No reference phase enters the reconstruction.
\end{paperalgorithm}

\FloatBarrier
\section{Complete reconstruction time and accuracy}\label{M-sec:runtime}
The complete-pipeline comparison holds local reconstruction and reconciliation fixed, and gives the grid every applicable implementation improvement. It therefore tests whether adaptive subdivision itself saves time.

\subsection{Inputs, controls, and measurements}\label{M-sec:timing}
The dataset contains 208 entries from six image families (Table~\ref{M-tab:corpus}). Ten inputs covering smooth fields, dense texture, and non-square dimensions guide implementation choices, while the other 198 evaluate the selected configurations. Because the dataset includes correlated cases and had already been used in exploratory work, this larger comparison evaluates fixed configurations rather than providing an independent test of generalization.

The exploratory subset comprises five photograph-derived inputs (trees, a portrait, noisy baboon, Lena, and a Barbara crop), two generated surfaces (Zernike and peaks), one synthetic InSAR field, one experimental holographic input, and one MRI simulation. Their dimensions range from $126\times134$ to $1024\times1024$ pixels. The complete dataset also includes narrow rectangular test fields and non-power-of-two dimensions, so the comparison exercises both the BRiDCT and fallback paths.

\begin{table}[!htbp]
\centering\small
\caption{Composition of the 208-entry dataset. Counts denote test inputs, not independent acquisitions. Ten entries guide implementation choices; the other 198 provide the main comparison. Reference phases comprise 158 known generating fields, 30 terrain-model references, and 20 algorithmic references. Agreement with the last category does not establish agreement with independent ground truth.}\label{M-tab:corpus}
\begin{tabular}{lrp{.68\textwidth}}
\toprule
Family & Entries & Source and reference \\
\midrule
Photographs & 45 & Image intensities mapped to known phase fields, including USC-SIPI images. \\
InSAR & 69 & 19 generated fields; 50 InSAR-DLPU entries, including 30 with terrain-model references \cite{zhou2024insardlpu}. \\
Synthetic surfaces & 18 & Analytic or generated fields with known phase. \\
Published tests & 4 & Ghiglia--Pritt images and supplied reference arrays \cite{ghiglia1998two}. \\
Holography & 40 & 20 synthetic and 20 experimental entries; the latter use supplied algorithmic wrap-count references \cite{gontarz2023phase}. \\
MRI simulations & 32 & Magnetic resonance imaging (MRI) simulations from the quantitative susceptibility mapping (QSM) challenge, with known phase and brain masks \cite{marques2021qsm}. \\
\bottomrule
\end{tabular}
\end{table}

The reconstruction timings use one thread on an Apple M3 Max. For these experiments, observations are encoded in 256 intensity levels per cycle, and local and assembled solutions are quantized to the same phase step. Non-square inputs are padded with zero intensity to a square of side $\max(H,W)$ for every method, where $H$ and $W$ are the original height and width. The whole padded domain is reconstructed; output is cropped before evaluation, and MRI accuracy uses the supplied brain mask. These choices are shared controls, but they limit direct extrapolation to floating-point observations or reconstruction only within a mask.

The timer surrounds the complete C++ reconstruction call, including partition construction, local solves, seam processing, output assembly, and temporary-buffer destruction. Input loading, encoding, process startup, and accuracy evaluation are excluded. Transform plans are prepared during an untimed warm-up. Seven randomly ordered repetitions are recorded for every image and configuration at minimum sides $s=8$ and $s=16$, without an adaptive leaf-budget limit. Grids use nominal side $s$, with partial tiles at image edges. First-call latency is evaluated separately.

For each image and method we take the median repeated time, then form the adaptive/grid ratio. Table~\ref{M-tab:final-optimization} reports the median of those paired ratios and the ratio of summed case times, which gives greater weight to expensive images. A time ratio of 1.25 therefore means 25\% slower reconstruction.

Accuracy is evaluated on the original image, or the supplied MRI mask. We subtract the mean reconstruction-minus-reference difference over those pixels to remove one common additive offset before computing the fraction of evaluated pixels with absolute error below $\pi$, denoted $\SP$, and the root-mean-square error (RMSE). Mean paired changes in these measures show whether a difference in speed is accompanied by a difference in reference agreement. For example, a success change of $-1$ percentage point means a smaller fraction of pixels within tolerance.

\subsection{Implementation optimizations and the grid control}\label{M-sec:optimized}
We reduce the cost of constructing the tree, processing its interfaces, and solving its tiles. Tree construction is specific to the adaptive methods; improvements to local solves and reconciliation are also given to the grid.

For unrestricted residue refinement, depth-first traversal replaces the priority queue because processing order no longer affects the final leaves. Residue computation and cumulative summation are combined in one pass, avoiding a separate residue-map pass. Since the input is encoded on 256 levels, residue evaluation uses integer phase differences. A lookup table handles differences of exactly half a cycle consistently with the reference wrapping convention.

For reconciliation, we collect each shared boundary as a contiguous interface. Finding its median and MAD requires selecting central values rather than sorting every sample. A disjoint-set structure tracks the relative offsets of tile groups as they are joined, avoiding a separate propagation traversal. Finally, FFTW replaces the slow direct-transform fallback on shapes unsupported by BRiDCT.

To measure the benefit of these changes, we compare with a quadtree implementation that uses a priority queue, ordered-map seam collection, and direct rectangular fallbacks. Both implementations already use BRiDCT on supported shapes. The changes reduce median paired runtime by about 51\% at $s=8$ and 42\% at $s=16$. The comparison with the grid below therefore evaluates adaptivity after these avoidable costs have been reduced.

Two further shortcuts also fail to reduce paired runtime on the ten exploratory inputs. Direct integration with an explicit consistency check is about 1--4\% slower than DCT solves, while searching each tile only until its first residue is found is about 3--4\% slower than building the fused residue table. The retained pipeline therefore keeps DCT solves and the full table.

\subsection{Measured outcome}\label{M-sec:results}
\begin{figure}[!htbp]
\centering\includegraphics[width=\textwidth]{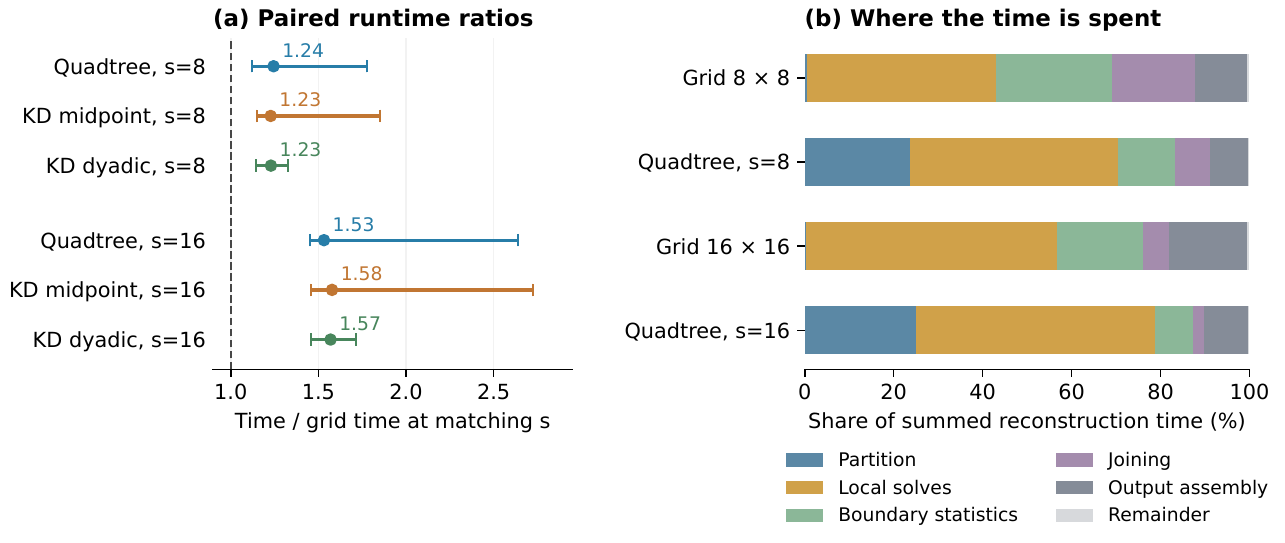}
\caption{Complete reconstruction time relative to the grid and its stage breakdown. (a) Each adaptive method uses minimum tile side $s$ and is paired with a grid of nominal side $s$ on the same input. Per-entry times are medians over seven repeats. Dots show the median ratio across 198 entries; whiskers show the 10th--90th percentiles across entries, not confidence intervals. The dashed line at one marks equal time. (b) Stage times are summed across entries using their per-entry medians and divided by the summed median complete-call times. Partitioning includes residue computation and tree construction; local solves include data preparation, the Poisson right-hand side, and transforms. Boundary statistics compute seam medians and MAD; joining propagates offsets. The remainder includes cleanup, uninstrumented overhead, and differences caused by summarizing each stage separately.}\label{M-fig:runtime-comparison}
\end{figure}

The optimized grid remains faster (Table~\ref{M-tab:final-optimization}). Quadtree reconstruction takes 24\% longer in median paired time at $s=8$, and 53\% longer at $s=16$. Both kd-tree variants also remain slower than the grid. Across the 198 entries, no tested adaptive configuration has a lower per-image median time than the optimized grid at either minimum size. Figure~\ref{M-fig:runtime-comparison} shows the spread across inputs and the contribution of each stage.

\begin{table}[!htbp]
\centering\small
\caption{Complete reconstruction time and reference agreement on 198 entries. $s$ is the minimum adaptive tile side in pixels; the grid uses nominal $s\times s$ tiles. For each entry, $t$ and $t_G$ are the median times over seven repeats for the adaptive method and its grid control. The median of $t/t_G$ describes a typical entry, while $\sum t/\sum t_G$ compares aggregate time across entries. Ratios above one mean slower reconstruction. $\Delta\SP$ and $\Delta$RMSE are mean paired changes, adaptive minus grid, in percentage points (pp) and radians. $\SP$ is the fraction of pixels with absolute error below $\pi$ after removing a common mean phase offset. Thus positive $\Delta\SP$ and negative $\Delta$RMSE indicate better reference agreement. All methods share BRiDCT, FFTW fallback, and reconciliation.}\label{M-tab:final-optimization}
\begin{tabular}{llrrrr}
\toprule
$s$ (px) & Partition & Median $t/t_G$ & $\sum t/\sum t_G$ & $\Delta\SP$ (pp) & $\Delta$RMSE (rad) \\
\midrule
8 & Quadtree & 1.244 & 1.371 & -0.20 & +0.105 \\
8 & KD midpoint & 1.228 & 1.381 & -0.04 & +0.133 \\
8 & KD dyadic & 1.229 & 1.294 & -0.06 & +0.145 \\
\addlinespace
16 & Quadtree & 1.532 & 1.777 & -1.02 & +0.184 \\
16 & KD midpoint & 1.578 & 1.796 & -0.99 & +0.213 \\
16 & KD dyadic & 1.570 & 1.644 & -1.01 & +0.225 \\
\bottomrule
\end{tabular}

\end{table}

The two runtime summaries reveal different effects of binary subdivision. At $s=8$, both binary variants are close to the quadtree in median runtime; at $s=16$, neither improves that statistic. Dyadic cuts reduce summed time relative to midpoint cuts more than they change the median ratio, suggesting that their benefit is concentrated among expensive inputs. This advantage over midpoint cuts remains insufficient to beat the grid. Binary subdivision also has a construction cost: for the same number of leaves $L$, it requires $L-1$ splits, compared with $(L-1)/3$ for a full quadtree. Greater directional freedom therefore comes with additional node-processing work.

Balancing residues across a cut does not improve this result. On the ten exploratory inputs, the residue-balanced rule takes about 1.53 and 1.85 times the midpoint kd-tree runtime at $s=8$ and $s=16$, respectively, in median paired time with otherwise matched reconstruction. It also reduces mean success at both sizes. The broader comparison therefore retains the simpler midpoint and dyadic rules.

The additional runtime is not accompanied by a consistent improvement in reference agreement. Mean success changes are modest, but averages hide severe failures: two related shear-field test images fall from $\SP=100\%$ with the grid to $\SP=0\%$ with each adaptive method. Thus, after removing the common phase offset, none of their adaptive pixels lies within $\pi$ of the reference. Mean RMSE increases for all three adaptive configurations. The shear failures also occur with the quadtree before the construction optimizations, so they are not introduced by the faster construction routine. These observations do not prove that every adaptive reconstruction is worse; they rule out a claim that the tested allocation preserves accuracy across the dataset.

\subsection{Why fewer tiles do not save time}\label{M-app:cost}
Partition construction accounts for about 24--25\% of summed quadtree time, and local reconstruction for 47--54\% (Fig.~\ref{M-fig:runtime-comparison}). A regular grid avoids image-dependent partitioning. Adaptive tiling reduces the number of interfaces, but every pixel still contributes to a local solve, and retaining larger tiles can increase the work per pixel. For a fast transform on $n_t$ pixels, the work scales as $O(n_t\log n_t)$. Combining small tiles increases the transform size while leaving the total number of pixels unchanged, so fewer local problems need not mean less transform work.

The complexity follows the same separation of work. Let $N$ denote the processed pixel count, $L$ the number of leaves, $P$ the number of pixel pairs crossing tile boundaries, and $E$ the number of adjacent tile pairs. Residue preprocessing costs $O(N)$, after which the unrestricted traversal creates $O(L)$ nodes and sorts leaves into a deterministic order in $O(L\log L)$. The transform work over all tiles is $O(\sum_t n_t\log n_t)$. Replacing a direct separable transform on an $h\times w$ tile, whose cost is $O(hw(h+w))$, by a fast DCT avoids an additional penalty on unsupported rectangular shapes.

Fewer tiles mainly help the boundary stages: seam processing is linear in $P$ on average for the selection routine used, and reliability sorting costs $O(E\log E)$. Pixel preparation and output assembly still require $O(N)$ work, while storage is $O(N+E+L)$ apart from cached transform plans. Thus reducing interfaces leaves both the initial residue pass and the work over all pixels in place.

This distribution of work limits what further DCT acceleration can achieve on its own. The local-solve stage includes input extraction and right-hand-side construction as well as the transforms themselves. A faster DCT changes only part of that stage, and the same improvement is available to the grid. A useful optimization must therefore be judged by the complete paired runtime, rather than a kernel speedup in isolation.

For a single reconstruction, initialization adds a cost absent from the warm timings. On the ten exploratory inputs, median first-call/warm-call ratios are about 2.2--2.9. Quadtree and dyadic kd-tree remain slower than the grid in paired first-call time; process startup, loading, and encoding are excluded.

\section{Subdivision criteria and reconstruction accuracy}\label{M-sec:criteria}
The runtime comparison fixes residue count as the subdivision criterion. We now compare nine criteria at two minimum tile sizes and two leaf budgets on all 208 entries. This separate study measures allocation and reference error under numerical conditions held fixed across criteria. It does not impose an error tolerance or rank the criteria by complete runtime.

\subsection{Stopping rules and priority scores}
A criterion assigns a nonnegative score to each cell, and a tile sums scores over its interior cells. The resulting score serves two purposes: zero stops refinement, while positive values determine priority when a leaf budget limits subdivision. Consequently, a sparse score can leave large zero-score regions untouched, whereas a score that is positive everywhere continues refining even where the local fit is already consistent.

\begin{figure}[!htbp]
\centering\includegraphics[width=.87\textwidth]{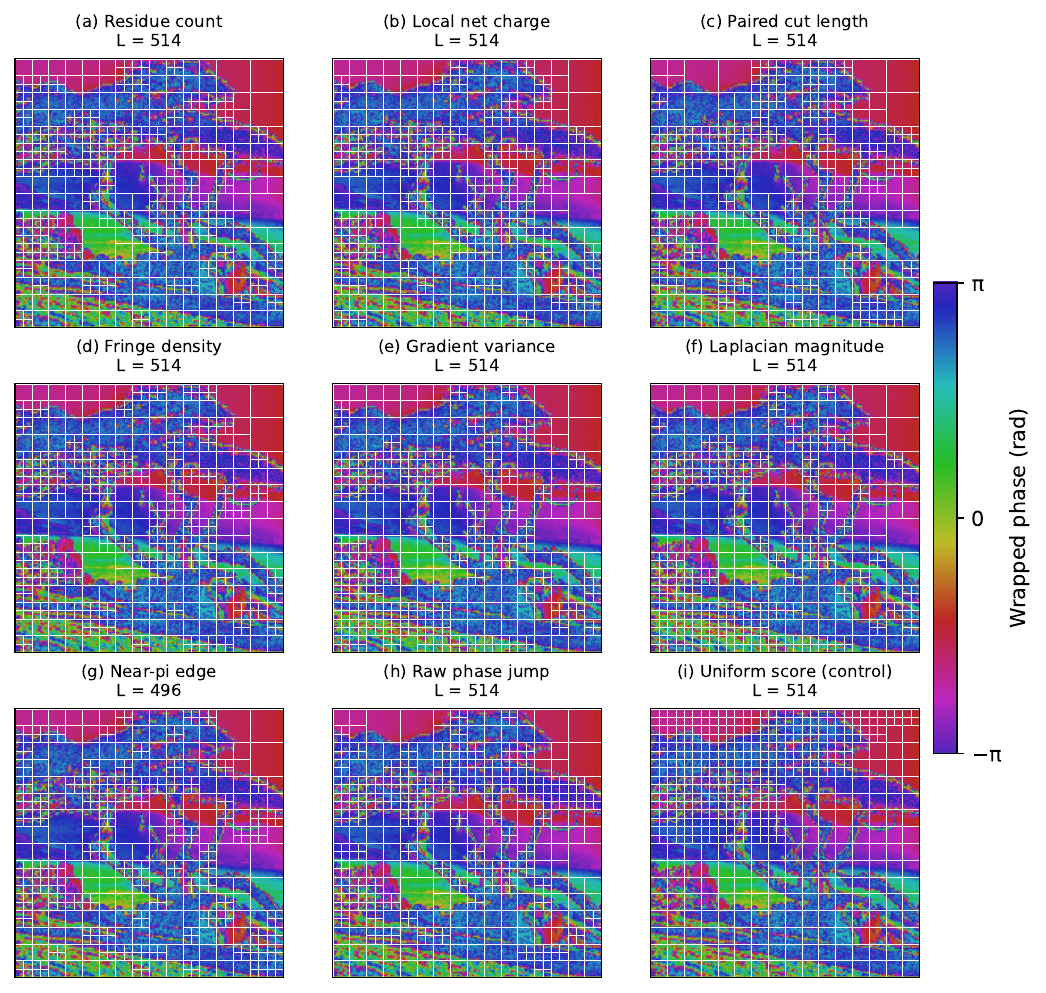}
\caption{Effect of the subdivision criterion on a fixed observation. All panels show the same unquantized $256\times256$ tree-photograph phase field as row B of Fig.~\ref{M-fig:overview}, with white quadtree boundaries over the wrapped phase. Panels (a)--(i) use criteria A--I from Sec.~\ref{M-sec:criteria}; $L$ is the resulting number of tiles. Minimum side $s=8$ pixels and stopping budget $B=512$ leaves are identical throughout. A four-way split can exceed this budget by two leaves. The shared cyclic color scale gives phase in radians. These partitions illustrate how scores allocate a limited tile budget; they are not the unrestricted partitions used in the main runtime comparison.}\label{M-fig:criteria-partitions}
\end{figure}

With fixed split locations and no budget limit, the stopping test determines the leaves; positive priorities affect only processing order. A restricted budget makes those priorities matter, as Fig.~\ref{M-fig:criteria-partitions} illustrates. Criteria A--C use residues, D--F describe phase variation, G--H test phase jumps, and I ignores image content. Their neighborhoods and thresholds are fixed across entries, without a claim that these settings are optimal.

\noindent\textbf{A. Residue count.} The score counts absolute residue charge, directing refinement toward differences that cannot be fitted consistently. This is the criterion used in the main timing comparison.

\noindent\textbf{B. Local net charge.} The score is the magnitude of the signed residue sum over an $8\times8$ cell neighborhood, with zeros outside the image. Opposite charges can cancel, so zero net charge does not imply absence of residues.

\noindent\textbf{C. Paired cut length.} This criterion pairs nearby opposite residues greedily within a Manhattan distance of 16 cells, prioritizing short connections, and connects unmatched residues to the nearest border. Each path contributes one to every cell it visits; overlapping paths accumulate. These paths guide tile allocation only; they do not delete edges from the local least-squares problem.

\noindent\textbf{D. Fringe density.} The score counts cell edges whose principal difference has magnitude above $3\pi/4$, emphasizing rapid phase variation. A steep, consistently sampled ramp can score highly without any residue.

\noindent\textbf{E. Gradient variance.} This criterion sums the horizontal and vertical principal-difference variances in $5\times5$ windows, with reflection at image boundaries, and averages the edge scores onto cells. It responds to texture, curvature, and noise rather than inconsistency alone.

\noindent\textbf{F. Laplacian magnitude.} The score averages the absolute divergence of principal differences over a cell's four corners, using replicated values at image boundaries. Because smooth curvature also produces a nonzero score, this criterion may subdivide regions whose observed differences are already consistent.

\noindent\textbf{G. Near-$\pi$ edge.} The score is one if any edge has principal-difference magnitude greater than $\pi-10^{-6}$ radians, and zero otherwise. This narrow test can miss inconsistencies with less extreme edge differences.

\Needspace{4\baselineskip}
\noindent\textbf{H. Raw phase jump.} This criterion marks neighboring wrapped values whose difference exceeds $\pi$ before rewrapping. It uses Baldi's subdivision test \cite{baldi2001twodimensional}, without reproducing his complete method. Smooth phase crossing a wrap boundary can activate it.

\noindent\textbf{I. Uniform score (control).} Every cell receives the same positive score. The tile score is its number of interior cells, so subdivision is driven by tile size and the budget, independently of image content.

\subsection{Criterion and budget evaluation}\label{M-sec:criterion-evaluation}
For each entry, we cross the nine criteria with minimum sides $s\in\{8,16\}$ pixels and the full and half budgets defined in Table~\ref{M-tab:criterion-evaluation}. This gives 7488 reconstructions.

For this allocation and accuracy study, every criterion uses the same double-precision DCT solver, local and final quantization, and median/MAD reliability-tree reconciliation. Scores and local right-hand sides use the original, unquantized observations without square padding. All nine criteria are compared under these same conditions. Their error differences describe the effect of allocation within this study and should not be combined numerically with the encoded-input BRiDCT results of Sec.~\ref{M-sec:runtime}.

\begin{table}[!htbp]
\centering\small
\caption{Criterion comparison on 208 entries at minimum side $s$ pixels. For original dimensions $H\times W$, the full budget is $B_1=\max(1,\lfloor H/s\rfloor\lfloor W/s\rfloor)$ and the half budget is $B_{1/2}=\max(1,\lfloor B_1/2\rfloor)$. For a $256\times256$ image at $s=8$, these targets are 1024 and 512 tiles; a split may exceed the target by two. $L$ is the median tile count. $\Delta\SP$ (percentage points, pp) and $\Delta$RMSE (radians, rad) are mean paired changes relative to residue count at the same $s$ and budget. Positive $\Delta\SP$ and negative $\Delta$RMSE improve agreement. The zero control rows do not imply equal accuracy between budgets: absolute control means are given below. Budgets limit tiles, not error, and this table does not compare runtimes.}\label{M-tab:criterion-evaluation}
\begin{tabular}{lrrrrrr}
\toprule
 & \multicolumn{3}{c}{Full budget $B_1$} & \multicolumn{3}{c}{Half budget $B_{1/2}$} \\
\cmidrule(lr){2-4}\cmidrule(lr){5-7}
Criterion & $L$ & $\Delta\SP$ (pp) & $\Delta$RMSE (rad) & $L$ & $\Delta\SP$ (pp) & $\Delta$RMSE (rad) \\
\midrule
\multicolumn{7}{l}{Minimum side $s=8$ pixels} \\
\midrule
A. Residue count & 302.5 & +0.00 & +0.000 & 302.5 & +0.00 & +0.000 \\
B. Local net charge & 419.5 & -0.25 & +0.266 & 419.5 & +0.50 & +0.236 \\
C. Paired cut length & 362.5 & -0.42 & +0.026 & 362.5 & -0.18 & +0.070 \\
D. Fringe density & 748 & +0.07 & +0.021 & 514 & +0.37 & -0.226 \\
E. Gradient variance & 1024 & -1.50 & +0.356 & 514 & +0.48 & +0.288 \\
F. Laplacian magnitude & 1024 & -1.53 & +0.356 & 514 & -0.38 & +0.350 \\
G. Near-$\pi$ edge & 1 & -5.50 & +0.216 & 1 & -4.79 & +0.179 \\
H. Raw phase jump & 841 & -0.72 & +0.163 & 514 & +0.07 & +0.187 \\
I. Uniform control & 1024 & -1.53 & +0.356 & 514 & -0.21 & +0.267 \\
\midrule
\multicolumn{7}{l}{Minimum side $s=16$ pixels} \\
\midrule
A. Residue count & 139 & +0.00 & +0.000 & 130 & +0.00 & +0.000 \\
B. Local net charge & 160 & +0.00 & +0.249 & 130 & +0.06 & +0.251 \\
C. Paired cut length & 146.5 & -0.17 & +0.013 & 130 & -1.22 & +0.167 \\
D. Fringe density & 247 & +0.09 & -0.021 & 130 & +0.42 & -0.063 \\
E. Gradient variance & 256 & -0.11 & +0.240 & 130 & +0.63 & +0.230 \\
F. Laplacian magnitude & 256 & -0.11 & +0.240 & 130 & -0.33 & +0.256 \\
G. Near-$\pi$ edge & 1 & -2.59 & -0.088 & 1 & -2.33 & +0.039 \\
H. Raw phase jump & 245.5 & +0.16 & +0.118 & 130 & -0.17 & +0.192 \\
I. Uniform control & 256 & -0.11 & +0.240 & 130 & -0.86 & +0.258 \\
\bottomrule
\end{tabular}

\par\smallskip\begin{minipage}{\textwidth}\footnotesize
\textit{Absolute residue-count control.} Values are mean $\SP$ (\%) / mean RMSE (rad), in the order $B_1$, $B_{1/2}$.
$s=8$: 84.628 / 4.336, 84.062 / 4.374.
$s=16$: 82.824 / 4.619, 82.275 / 4.500.

\end{minipage}
\end{table}

The two error measures in Table~\ref{M-tab:criterion-evaluation} need not improve together: more pixels can fall below the $\pi$ threshold while the largest errors increase. Nor do equal budgets imply equal tile counts, because a sparse criterion may stop before using its budget.

The absolute control values show that changing the budget can affect the two error measures differently. At $s=16$, halving the budget lowers mean success but also lowers mean RMSE. A reduction in one error statistic therefore does not establish that accuracy is preserved.

Dense scores often spend more tiles without improving reconstruction. At $s=8$ and budget $B_1$, residue count produces a median of 302.5 tiles, compared with 1024 for gradient variance, Laplacian magnitude, and the uniform control. Their mean success is about 1.50--1.53 percentage points lower, and mean RMSE is about 0.36 radians higher. Conversely, stopping almost everywhere is not sufficient: the near-$\pi$ criterion has a median of one tile but loses 5.50 percentage points of success. Tile reduction must therefore be assessed together with error.

Residue count is not the most accurate criterion in every setting. At $s=8$ and budget $B_{1/2}$, fringe density improves mean success by 0.37 percentage points and reduces mean RMSE by 0.23 radians, while its median tile count rises from 302.5 to 514. At $s=16$ with the same budget fraction, fringe density also improves both error measures, although both median counts are 130. Local net charge and gradient variance sometimes improve success while increasing RMSE. These findings support residue count as an interpretable baseline that often uses fewer tiles, rather than a universal accuracy optimum.

\Needspace{14\baselineskip}
\subsection{Relaxed stopping and runtime}
The criterion comparison shows how allocation affects error, but does not establish whether coarser partitions save time. We therefore test relaxed residue stopping on the ten exploratory inputs, measuring whether allowing a few residues inside a tile reduces reconstruction time without losing reference agreement. The density rules stop a tile when its residue count per interior cell is at most $0.001$ or $0.01$. A second rule stops when the count is below $0.25(h+w)$ for tile dimensions $h,w$, provided the larger side is at most $4s$. Each is paired with zero-residue stopping at $s=8$ and $s=16$, with five repeated timings per input. Both arms use the same joining rule and permit direct integration in place of a DCT solve only after checking consistency against every internal edge. These timings exclude temporary-buffer destruction and are interpreted only within this comparison.

These tests do not yield a consistent time--accuracy improvement. Across the two minimum sizes, the density rules take 1.6--12.4\% longer in median paired time and reduce mean success by 0.78--5.93 percentage points. The perimeter-based cutoff is about 0.9\% faster at $s=8$ and 2.3\% slower at $s=16$, while mean RMSE increases at both sizes. Allowing additional local inconsistency therefore does not by itself produce a useful speedup in these tests. A broader study should instead measure the fastest reconstruction available at each acceptable error level.

\section{Discussion and conclusions}
The optimized grid remains faster than the tested quadtree and kd-tree configurations, even after substantial reductions in adaptive reconstruction time. Stage measurements explain the result: residue preprocessing and larger local solves outweigh the savings at tile boundaries. The criterion study shows why refinement must also be assessed through accuracy: allocating more tiles does not consistently improve reference agreement, and gains depend on the score and budget. The value of an adaptive partition therefore depends on both its complete runtime and the reconstruction error it produces.

The runtime conclusion concerns single-threaded, encoded-input reconstruction on one machine, with unrestricted residue refinement. Related dataset entries and different reference types limit generalization to independent physical acquisitions. Padding, quantization, and reconciliation are also part of the evaluated pipeline. The comparison therefore establishes a result for these configurations with their common BRiDCT solver and FFTW fallback, rather than a universal ranking of adaptive methods.

These results leave open whether limited refinement can provide a faster approximate reconstruction when a larger error is acceptable. A targeted study would vary the criterion, tile budget, and minimum tile size for the adaptive methods, and the tile size for the grid. For each setting, it would measure complete reconstruction time and error against the reference phase. At each allowed error level, the comparison would identify the fastest setting that meets that tolerance. This would show whether the quadtree or kd-tree offers an advantage for an approximate first result, and how that advantage changes as the required accuracy increases.

The kd-tree comparison could also be extended with synthetic images whose difficult regions have controlled shapes. For example, residues could be concentrated in a compact patch or a narrow band, while image size, noise level, and residue count are held fixed. Comparing quadtree and kd-tree reconstruction times at similar reference error would then test whether binary cuts help specifically with elongated structures. This would isolate a geometric effect that the mixed-dataset comparison does not resolve, while retaining the shared solver and joining rule.

Further engineering could reduce repeated gradient preparation or group tiles of equal shape to improve data reuse. Transform plans and buffers are already cached; additional gains from these proposals remain to be demonstrated. The report's practical conclusion is to retain the grid as the runtime baseline and require any adaptive improvement to survive a complete reconstruction comparison. Fewer tiles are a useful geometric property; they become a computational advantage only when the work they remove exceeds the work required to construct and process them.

\section*{Data and code availability}
Public source collections are identified in Table~\ref{M-tab:corpus}. The photograph-derived inputs include images attributed to the USC-SIPI collection.\footnote{USC-SIPI Image Database: \url{https://sipi.usc.edu/database/}.} The assembled derived arrays, case lists, preprocessing scripts, experiment code, and outputs are retained by the authors and are not currently publicly available. Some locally prepared inputs lack a complete generation recipe, so citations to the source collections do not reproduce the exact benchmark.

\section*{Supplementary material}
The supplementary material contains the detailed input inventory, additional mathematical and implementation details, the underlying criterion score maps, a separate baseline DCT-library comparison, and expanded results for the exploratory variants and first-call measurements.

\begingroup
\footnotesize
\setlength{\parskip}{0pt}
\providecommand{\BIBdecl}{\setlength{\itemsep}{0pt}}
\bibliographystyle{IEEEtran}
\bibliography{refs}

@article{antonopoulos2015tilebased,
  author  = {Antonopoulos, Georgios C. and Steltner, Benjamin and
             Heisterkamp, Alexander and Ripken, Tammo and Meyer, Heiko},
  title   = {Tile-Based Two-Dimensional Phase Unwrapping for Digital
             Holography Using a Modular Framework},
  journal = {PLOS ONE},
  volume  = {10}, number = {11}, pages = {e0143186}, year = {2015},
  doi     = {10.1371/journal.pone.0143186}
}

@article{baldi2001twodimensional,
  author = {Baldi, Antonio},
  title = {Two-dimensional phase unwrapping by quad-tree decomposition},
  journal = {Applied Optics},
  year = {2001},
  volume = {40},
  number = {8},
  pages = {1187--1194},
  doi = {10.1364/AO.40.001187}
}

@article{ghiglia1994robust,
  author  = {Ghiglia, Dennis C. and Romero, Louis A.},
  title   = {Robust Two-Dimensional Weighted and Unweighted Phase Unwrapping
             That Uses Fast Transforms and Iterative Methods},
  journal = {Journal of the Optical Society of America A},
  volume  = {11}, number = {1}, pages = {107--117}, year = {1994},
  doi     = {10.1364/JOSAA.11.000107}
}

@book{ghiglia1998two,
  author    = {Ghiglia, Dennis C. and Pritt, Mark D.},
  title     = {Two-Dimensional Phase Unwrapping: Theory, Algorithms, and
               Software},
  publisher = {Wiley-Interscience}, year = {1998}
}

@article{itoh1982analysis,
  author  = {Itoh, Kazuyoshi},
  title   = {Analysis of the Phase Unwrapping Algorithm},
  journal = {Applied Optics},
  volume  = {21}, number = {14}, pages = {2470}, year = {1982},
  doi     = {10.1364/AO.21.002470}
}

@misc{moevus2026tilebased,
  author        = {Moevus, Antoine and Mignotte, Max},
  title         = {Translation-Invariant Tile-Based Phase Unwrapping with Residual-Weighted Multipath Averaging},
  year          = {2026},
  eprint        = {2609.13409},
  archivePrefix = {arXiv},
  primaryClass  = {eess.IV},
  note          = {arXiv:2609.13409},
  url           = {https://arxiv.org/abs/2609.13409}
}

@article{strand1999two,
  author = {Strand, J. and Taxt, T. and Jain, A.K.},
  title = {Two-dimensional phase unwrapping using a block least-squares method},
  journal = {IEEE Transactions on Image Processing},
  year = {1999},
  volume = {8},
  number = {3},
  pages = {375-386},
  doi = {10.1109/83.748892}
}

@article{yu2011residues,
  author = {Hanwen Yu and Zhenfang Li and Zheng Bao},
  title = {Residues Cluster-Based Segmentation and Outlier-Detection Method for Large-Scale Phase Unwrapping},
  journal = {IEEE Transactions on Image Processing},
  year = {2011},
  volume = {20},
  number = {10},
  pages = {2865-2875},
  doi = {10.1109/TIP.2011.2138148}
}

@article{zhou2024insardlpu,
 author={Zhou, Lifan and Yu, Hanwen},
 title={{InSAR-DLPU}: A Benchmark Dataset for Deep Learning-Based {SAR} Interferometry Phase Unwrapping},
 journal={IEEE Geoscience and Remote Sensing Magazine},
 volume={12}, number={2}, pages={118--124}, year={2024}, doi={10.1109/MGRS.2024.3359691}
}

@article{gontarz2023phase,
  author={Gontarz, Micha{\l} and Dutta, Vibekananda and Kujawi{\'n}ska, Ma{\l}gorzata and Krauze, Wojciech},
  title={Phase unwrapping using deep learning in holographic tomography},
  journal={Optics Express},volume={31},number={12},pages={18964--18992},year={2023},doi={10.1364/OE.486984}
}

@article{marques2021qsm,
 author={Marques, Jos{\'e} P. and Meineke, Jakob and Milovic, Carlos and Bilgic, Berkin and Chan, Kwok-Shing and Hedouin, Renaud and van der Zwaag, Wietske and Langkammer, Christian and Schweser, Ferdinand},
 title={{QSM} reconstruction challenge 2.0: A realistic in silico head phantom for {MRI} data simulation and evaluation of susceptibility mapping procedures},
 journal={Magnetic Resonance in Medicine},volume={86},number={1},pages={526--542},year={2021},doi={10.1002/mrm.28716}
}

@article{shao2008typeii,
 author={Xuancheng Shao and Steven G. Johnson},
 title={Type-{II}/{III} {DCT}/{DST} algorithms with reduced number of arithmetic operations},
 journal={Signal Processing}, volume={88}, number={6}, pages={1553--1564}, year={2008},
 doi={10.1016/j.sigpro.2008.01.004}
}

@misc{ooura2001general,
 author={Takuya Ooura}, title={General Purpose FFT (Fast Fourier/Cosine/Sine Transform) Package},
 year={2001}, howpublished={Software package},
 url={https://www.kurims.kyoto-u.ac.jp/~ooura/fft.html}
}

@article{chen2002phase,
  author = {Chen, Curtis W. and Zebker, Howard A.},
  title = {Phase Unwrapping for Large {SAR} Interferograms: Statistical Segmentation and Generalized Network Models},
  journal = {IEEE Transactions on Geoscience and Remote Sensing},
  year = {2002}, volume = {40}, number = {8}, pages = {1709--1719},
  doi = {10.1109/TGRS.2002.802453}
}

@article{moevus2026bridct,
  author = {Moevus, Antoine and Mignotte, Max},
  title = {{BRiDCT}: Fast Two-Dimensional {DCTs} Using {SIMD}: {SIMD} Organization, Register Blocking, and Numerical Verification},
  journal = {arXiv preprint arXiv:2609.28519},
  year = {2026},
  eprint = {2609.28519},
  archivePrefix = {arXiv},
  url = {https://arxiv.org/abs/2609.28519}
}
\endgroup

\clearpage
\setcounter{section}{0}
\setcounter{subsection}{0}
\setcounter{figure}{0}
\setcounter{table}{0}
\setcounter{equation}{0}
\setcounter{paperalgorithm}{0}
\setcounter{footnote}{0}
\setcounter{page}{1}
\renewcommand{\thepage}{S\arabic{page}}
\renewcommand{\thesection}{S\arabic{section}}
\renewcommand{\thefigure}{S\arabic{figure}}
\renewcommand{\thetable}{S\arabic{table}}
\renewcommand{\theequation}{S\arabic{equation}}
\renewcommand{\thepaperalgorithm}{S\arabic{paperalgorithm}}
\def\theHsection{S.\arabic{section}}
\def\theHsubsection{S.\arabic{section}.\arabic{subsection}}
\def\theHfigure{S.\arabic{figure}}
\def\theHtable{S.\arabic{table}}
\def\theHequation{S.\arabic{equation}}
\def\theHpaperalgorithm{S.\arabic{paperalgorithm}}
\def\theHfootnote{S.\arabic{footnote}}

\raggedbottom
\widowpenalty=10000\clubpenalty=10000
\begin{center}
{\Large Adaptive Tiling for Least-Squares Phase Unwrapping\par}
\vspace{.4em}{\large Runtime and Accuracy --- Supplementary Material\par}
\vspace{.8em}Antoine Moevus \quad Max Mignotte\\[.4em]
Universit\'e de Montr\'eal \quad September 2026
\end{center}
This supplement provides the detailed input descriptions, additional method analysis, score maps, and supporting measurements for the accompanying technical report. The primary result is the complete-reconstruction comparison using BRiDCT in the main report's Table~\ref{M-tab:final-optimization}. The baseline DCT-library comparison here evaluates a different implementation and is reported separately.
\section{Data and measurement details}\label{S-app:measurements}
The 208-entry dataset contains 158 entries with a known generating reference, 30 real radar-interferometry entries with terrain-model references, and 20 experimental holographic entries with algorithmic references (Table~\ref{S-tab:corpus}). Agreement with an algorithmic reference is not independent physical ground truth. MRI inputs use cropped quantitative susceptibility mapping (QSM) simulations \cite{marques2021qsm} and their supplied brain masks; other inputs use the full image. 

For the reconstruction-time experiments, the input is first encoded in 256 intensity levels per cycle. Non-square arrays are placed at the top left of a square of side $\max(H,W)$ and padded with zero intensity. Reconstruction covers this whole square, including padding and pixels outside evaluation masks. Output is cropped back to the original dimensions before evaluation. The dimensions in Tables~\ref{S-tab:corpus} and~\ref{S-tab:followup-cases} describe the original arrays, not the padded working domain.

\begin{table}[!htbp]
\centering\small
\caption{Dataset families, original image dimensions, and reference-phase construction. Dimensions are height $\times$ width in pixels before square padding; ranges cover the selected entries, not the entire source collections. Known generating references include both analytic fields and phases derived from photograph intensities. Terrain-model and algorithmic references are identified separately in the last column.}\label{S-tab:corpus}
\begin{tabular}{>{\raggedright\arraybackslash}p{.14\textwidth}r>{\raggedright\arraybackslash}p{.18\textwidth}>{\raggedright\arraybackslash}p{.50\textwidth}}
\toprule
Family & Entries & Dimensions & Source and reference phase \\
\midrule
Photograph-derived & 45 & \mbox{$256\times256$}, \mbox{$512\times512$}, \mbox{$1024\times1024$}; \mbox{$250\times288$} & Grayscale test images, including USC-SIPI images, scaled into phase fields; the generating field is the reference. Clean and noisy variants are included. \\
InSAR & 69 & \mbox{$256\times256$} & 19 locally generated interferograms with known phase; 20 simulated and 30 real TanDEM-X entries from InSAR-DLPU \cite{zhou2024insardlpu}. The real-data references derive from a digital elevation model. \\
Synthetic surfaces & 18 & \mbox{$256\times256$}, \mbox{$257\times257$}; \mbox{$458\times152$} & Analytic and generated phase fields, with the generating field retained as the reference. \\
Published test images & 4 & \mbox{$257\times257$}; \mbox{$458\times157$}; \mbox{$458\times152$} & Ghiglia--Pritt test images \cite{ghiglia1998two}, using their supplied reference arrays. \\
Holography & 40 & \mbox{$256\times256$} & 20 synthetic fields with known phase and 20 experimental inputs from Gontarz et al.\ \cite{gontarz2023phase}. Experimental references are reconstructed from the supplied quality-guided phase-unwrapping (QGPU) wrap counts. \\
MRI simulations & 32 & $92$--$167$ rows; $101$--$155$ columns & Cropped slices from four QSM challenge simulation/noise combinations \cite{marques2021qsm}, two echo times per view. Reference phase is $2\pi f\,\mathrm{TE}$ from the supplied frequency $f$ and echo time $\mathrm{TE}$. \\
\bottomrule
\end{tabular}

\end{table}\begin{table}[!htbp]
\centering\small
\caption{Ten inputs used for implementation choices and first-call measurements. Dimensions are original height $\times$ width in pixels, before padding. F1--F10 identify entries within this report. The last column states their origin and the image property relevant to these tests. Source-collection citations identify provenance; the exact derived benchmark arrays are not publicly available.}\label{S-tab:followup-cases}
\begin{tabular}{lllp{.46\textwidth}}
\toprule
ID & Input & Size & Origin and role in the comparison \\
\midrule
F1 & Tree photograph & $256\times256$ & Source recorded as USC-SIPI; intensity-derived phase with fine texture. \\
F2 & Portrait & $1024\times1024$ & Source recorded as USC-SIPI; larger intensity-derived phase image. \\
F3 & Zernike field & $256\times256$ & Locally generated polynomial phase field; curved fringes. \\
F4 & Peaks field & $256\times256$ & Locally generated smooth test surface; no residues in the input. \\
F5 & Noisy baboon & $512\times512$ & Locally prepared noisy photograph-derived field; dense subdivision. \\
F6 & Lena & $512\times512$ & Photograph-derived field retained in the local collection; intermediate subdivision. \\
F7 & Barbara crop & $250\times288$ & Locally cropped photograph-derived field; non-square leaves. \\
F8 & Synthetic InSAR & $256\times256$ & Locally generated interferometric field; known generating phase. \\
F9 & Holographic image & $256\times256$ & Gontarz et al.\ \cite{gontarz2023phase}, test sample 00189; algorithmic reference. \\
F10 & MRI simulation & $126\times134$ & QSM challenge \cite{marques2021qsm}, Sim1Snr1, axial slice 81, echo 3; cropped reference mask. \\
\bottomrule
\end{tabular}
\end{table}

For a valid-pixel set $M$, let $e_p=\widehat\phi_p-\phi_p$ and subtract its mean $\bar e$ to remove the arbitrary constant phase offset. The success fraction is $\SP=|M|^{-1}\sum_{p\in M}\mathbf1\{|e_p-\bar e|<\pi\}$; RMSE is $[|M|^{-1}\sum_{p\in M}(e_p-\bar e)^2]^{1/2}$. Success measures the fraction within tolerance, whereas RMSE is sensitive to large errors. The main report's Table~\ref{M-tab:final-optimization} reports mean paired changes in success in percentage points and RMSE in radians. Its median runtime statistic summarizes paired ratios; its summed-time statistic is explicitly a ratio of summed case times.

\section{Method specification}\label{S-app:method}
\subsection{Local fitting and residues}\label{S-sec:integrability}
Let $\phi$ be a phase field and $\psi=\W(\phi)$ its wrapped observation, where $\W$ reduces angles to $(-\pi,\pi]$. An oriented edge $e=(p,q)$ joins two neighboring pixels. On tile $T$, collect the observed principal differences $g_e=\W(\psi_q-\psi_p)$ into $g_T$. The operator $D_T$ computes the corresponding differences of a candidate phase, so $(D_Tu)_e=u_q-u_p$. The local problem is
\begin{equation}
 u_T\in\arg\min_u\|D_Tu-g_T\|_2^2.\label{S-eq:ls}
\end{equation}
Its normal equations, $D_T^{\mathsf T}D_Tu=D_T^{\mathsf T}g_T$, form a discrete Poisson system with the natural boundary conditions of the local fit. No edge outside the tile is included. The DCT diagonalizes this system \cite{ghiglia1994robust}. Its zero-frequency coefficient is fixed to zero because neighboring differences do not determine an additive constant. Local solutions are rounded to a phase step $\delta=2\pi/256$ radians, with ties to even. After offset assembly, the output minimum is subtracted and the values are rounded to the same step. Input encoding is used in the timing comparison, whereas the criterion study retains the original floating-point observations.

For a four-pixel cell $c$, let $r_c=(2\pi)^{-1}\sum_{e\in\partial c}\epsilon_{ce}g_e$, with signs following the oriented cell boundary. The residue criterion is
\begin{equation}
 C_r(T)=\sum_{c\subset T}|r_c|,\label{S-eq:count}
\end{equation}
using only cells wholly inside the tile. In exact arithmetic on a hole-free rectangle, $C_r(T)=0$ exactly when these differences admit a consistent potential: zero circulation on all cells makes path integration independent of the path. The unrounded least-squares residual is then zero, and the solution is unique up to a constant.

This concerns observed differences, not the true phase. Slopes $3\pi/2$ and $-\pi/2$ per pixel have identical wrapped observations and zero residues. Recovering the physical phase needs additional sampling assumptions \cite{itoh1982analysis}. Cells crossed by tile boundaries are also absent from the local count, leaving inconsistencies for reconciliation.

\subsection{Budget conventions}
The reconstruction and median/MAD reconciliation are defined in the main report, Sec.~\ref{M-sec:joining} and Algorithms~\ref{M-alg:partition}--\ref{M-alg:reconstruct}. In the criterion comparison, the size-based budget is $B_1=\max(1,\lfloor H/s\rfloor\lfloor W/s\rfloor)$; the restricted budget is $B_{1/2}=\max(1,\lfloor B_1/2\rfloor)$. A regular grid includes partial edge tiles and has $\lceil H/s\rceil\lceil W/s\rceil$ tiles. On non-dyadic images, bisection need not reproduce that grid.

\subsection{Complexity of the measured implementation}\label{S-app:cost}
Let $N$ be the processed pixel count, including padding, and $L$ the final tile count. Residue evaluation and a summed-area table cost $O(N)$ and give constant-time tile-score queries. Priority refinement costs $O(L\log L)$ in queue operations. The final unrestricted traversal instead creates $O(L)$ nodes, then sorts leaves into a deterministic order in $O(L\log L)$; it removes queue overhead without changing this overall bound. Fixed-neighborhood residue, fringe, variance, and divergence maps are linear in $N$, whereas paired-path criteria also require residue searches and path construction.

Fast-transform work on $n_t$ pixels scales as $O(n_t\log n_t)$, yielding $O(\sum_t n_t\log n_t)$ local work. A direct separable fallback on an $h_t\times w_t$ rectangle costs $O(h_tw_t(h_t+w_t))$; replacing it can dominate aggregate speedups on unusual shapes. Input extraction, right-hand-side construction, pixel ownership, and output assembly additionally require linear pixel work regardless of tile count.

Let $P$ count pixel pairs across tile boundaries, $E$ adjacent tile pairs, and $p_e$ the length of seam $e$. The final collector walks contiguous interfaces in $O(P)$ and computes median/MAD by selection, with average linear work in $p_e$ for the implementation used. Interfaces are first sorted by their tile identifiers to fix tie order, then stably sorted by decreasing reliability. These two sorts cost $O(E\log E)$. Weighted union--find propagates offsets while joining groups in $O(E\alpha(L))$ amortized work, where $\alpha$ is the inverse Ackermann function. The spatial and reconciliation storage is $O(N+E+L)$, excluding reusable transform plans. An ordered map over boundary samples would instead add $O(P\log(E+1))$ grouping work.

For the same leaf count $L$, a binary tree performs $L-1$ splits, compared with $(L-1)/3$ for a full quadtree. The tested residue-balanced cut uses binary searches of summed-area queries, adding $O(\log\ell_v)$ work at split $v$ along an axis of length $\ell_v$. The rejected early-exit search caps inspected cells at a fixed multiple of $N$ before building a prefix table, bounding repeated scanning but adding work before fallback. These bounds describe costs; the stage measurements determine their practical balance.

\subsection{Exact candidate score definitions}\label{S-app:score-definitions}
Tile scores sum cell scores $a_c$ over cells wholly inside the tile. These definitions instantiate the criteria in the main report, Sec.~\ref{M-sec:criteria}. Table~\ref{M-tab:criterion-evaluation} in the main report compares their allocation and accuracy. The complete-runtime comparison retains residue count (A), whose additional stopping variants are defined in Sec.~\ref{S-sec:alternatives}. Positive global rescaling changes neither priority order nor the zero test.

\noindent\textbf{A. Residue count.} $a_c=|r_c|$.

\noindent\textbf{B. Local net charge.} Absolute signed-residue sum in an $8\times8$ cell neighborhood, zero-padded outside the image.

\noindent\textbf{C. Paired cut length.} Greedily pair opposite residues within Manhattan radius 16, using up to eight candidates per direction and six rounds, shortest pairs first. Connect unmatched residues to the nearest border. Each rasterized segment adds one to visited cells; overlaps accumulate.

\noindent\textbf{D. Fringe density.} Count cell edges with $|\W(\psi_q-\psi_p)|>3\pi/4$.

\noindent\textbf{E. Gradient variance.} Sum horizontal and vertical principal-difference variances in reflected $5\times5$ windows; average edge-grid variances onto adjacent cells.

\noindent\textbf{F. Laplacian magnitude.} Take the divergence of principal differences with replicated image boundaries, then average its absolute value over the cell's four corners.

\noindent\textbf{G. Near-$\pi$ edge.} Set one if any cell edge has $|\W(\psi_q-\psi_p)|>\pi-10^{-6}$, zero otherwise.

\noindent\textbf{H. Raw phase jump.} Set one if any cell edge has $|\psi_q-\psi_p|>\pi$ before rewrapping, zero otherwise.

\noindent\textbf{I. Uniform score (control).}\label{S-crit:uniform} $a_c=1$, giving $C(T)=(h_T-1)(w_T-1)$ for an $h_T\times w_T$ tile. Equal-priority ties use fixed ordering.

\begin{figure}[!htbp]
\centering\includegraphics[width=.88\textwidth]{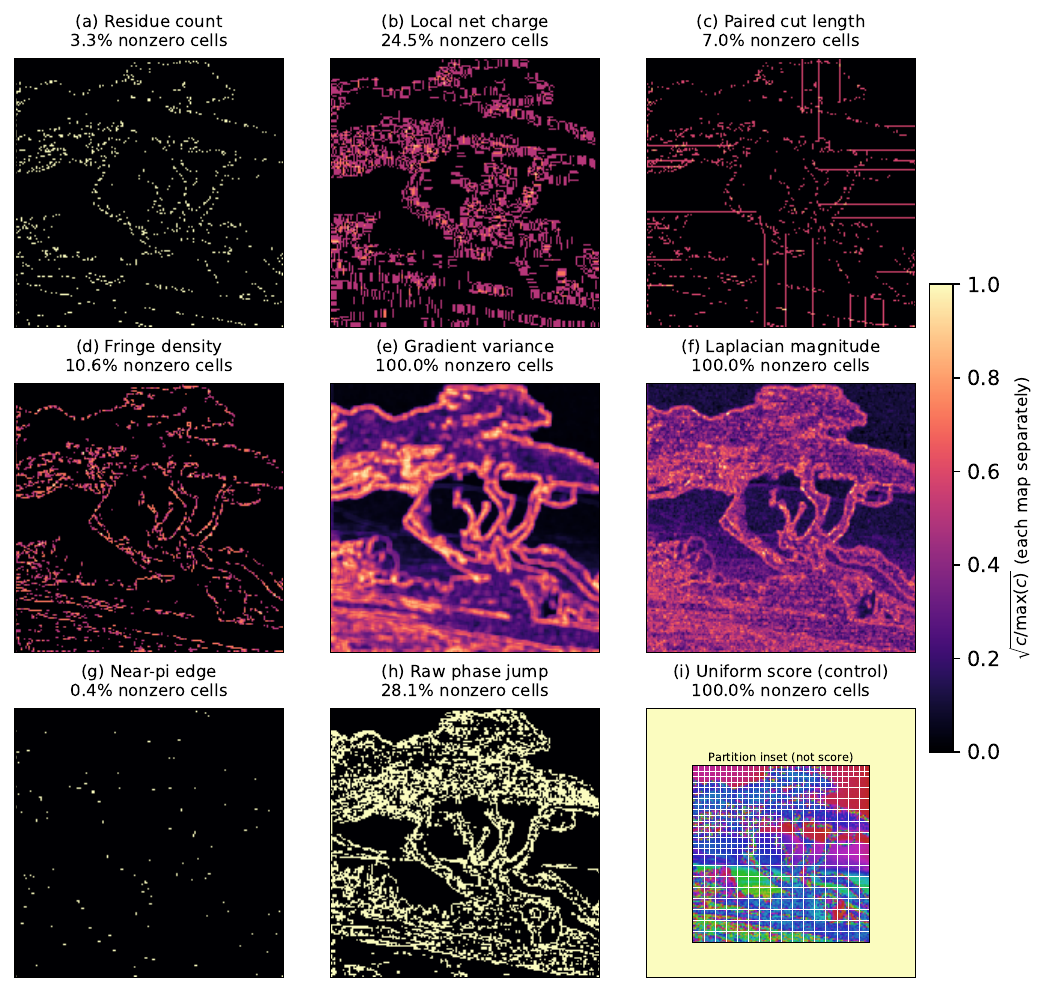}
\caption{Cell-score maps underlying the tree-image partitions in the main report's Fig.~\ref{M-fig:criteria-partitions}. Panels (a)--(i) correspond to criteria A--I. Color represents $\sqrt{a_c/a_{\max}}$, where $a_c$ is the cell score and $a_{\max}$ is the maximum within that panel; a zero map would be displayed as zero. Independent normalization reveals each map's spatial support but prevents comparison of absolute scores between criteria. Panel labels report the proportion or count of nonzero cells. The uniform control is one everywhere; its inset shows the resulting partition, not a second score map.}\label{S-fig:criteria-maps}
\end{figure}

Figure~\ref{S-fig:criteria-maps} shows the scores underlying the main report's partition comparison (Fig.~\ref{M-fig:criteria-partitions}). Positive rescaling leaves the partition rule unchanged, so each map is displayed with its own normalization. This visualization makes the difference between sparse and nearly everywhere-positive criteria visible without claiming a runtime or accuracy ranking.
\section{Baseline DCT-library comparison}\label{S-sec:dct-comparison}
A separate kernel benchmark compares a baseline Shao--Johnson implementation \cite{shao2008typeii}, before the BRiDCT optimizations, with FFTW 3.3.11 using measured plans, native Apple Accelerate, and float adaptations of Ooura's general and specialized routines \cite{ooura2001general}. Tests used an Apple M3 Max, one thread, single precision, and power-of-two square sides 8--256, the supported range of the square implementation tested here. We timed a normalized forward/inverse DCT pair and a Poisson spectral solve, which additionally divides by the Laplacian eigenvalues. Timing includes data rearrangement but excludes setup, allocation, and right-hand-side construction. Nine randomly ordered batches cycle through 16 inputs. All 225 numerical checks pass against double-precision references, with relative error below $4\times10^{-7}$ on nonconstant cases.

The library comparison in Table~\ref{S-tab:dct-libraries} uses FFTW's measured planning mode, which times candidate algorithms during setup.\footnote{FFTW planner flags: \url{https://www.fftw.org/fftw3_doc/Planner-Flags.html}.} Plans and working buffers are reused during timing. In this comparison, the Accelerate and Ooura routines are applied along rows and columns, including the data rearrangement needed for the two-dimensional transform. These routines are comparators in the library benchmark; the final reconstruction experiments use BRiDCT with the FFTW fallback.

The baseline Shao--Johnson implementation is fastest at sides 8--64 for both tasks (Table~\ref{S-tab:dct-libraries}). This size range is relevant to the partitions: more than 99\% of pooled quadtree leaves on the ten inputs in Table~\ref{S-tab:followup-cases} have both sides at most 64 pixels, at either minimum size. At $16\times16$, its spectral solve takes $0.189\,\mu$s versus $0.489\,\mu$s for specialized Ooura. At sides 128 and 256, it ranks second to Accelerate and takes about 1.5--1.6 times as long. These measurements support the choice of Shao--Johnson's factorization for small tiles, but do not measure BRiDCT's performance. The complete-reconstruction results in the main report, Sec.~\ref{M-sec:runtime}, include the BRiDCT optimizations and the common FFTW fallback.

\begin{table}[!htbp]
\centering\small
\caption{Poisson spectral-solve time in $\mu$s per square tile on an Apple M3 Max, using one thread and single precision. Each value is the median of nine randomly ordered timing batches cycling through 16 inputs. The operation comprises a forward DCT, division by nonzero Poisson eigenvalues, and an inverse DCT, including normalization and data rearrangement; setup, allocation, and right-hand-side construction are excluded. SJ baseline is the Shao--Johnson implementation preceding BRiDCT, not a BRiDCT measurement. Bold marks the fastest tested implementation at each size. Dashes indicate sizes unsupported by the tested interface, not by the entire algorithm family. Smaller values are faster; these kernel timings are not complete reconstruction times.}\label{S-tab:dct-libraries}
\begin{tabular}{rrrrrr}
\toprule
Side (px) & SJ baseline & FFTW & Accelerate & Ooura general & Ooura specialized \\
\midrule
8 & \textbf{0.048} & 0.131 & --- & 0.530 & 0.101 \\
16 & \textbf{0.189} & 1.189 & 0.956 & 1.712 & 0.489 \\
32 & \textbf{0.882} & 4.374 & 3.350 & 6.451 & --- \\
64 & \textbf{6.141} & 17.668 & 11.417 & 25.823 & --- \\
128 & 67.924 & 94.720 & \textbf{44.160} & 104.776 & --- \\
256 & 296.262 & 468.333 & \textbf{186.678} & 473.060 & --- \\
\bottomrule
\end{tabular}

\end{table}

\section{Additional implementation tests}\label{S-sec:alternatives}
The ten inputs in Table~\ref{S-tab:followup-cases} are used to compare bounded alternatives before the 198-entry evaluation. These exploratory tests select configurations; they do not rank criteria B--I. Comparisons are made within each timing session, since sessions were run separately on a shared machine. The primary reconstruction table uses a timer around the entire call, including temporary-buffer destruction. Early exploratory component timings excluded that cleanup and are not pooled with the primary measurements.

Table~\ref{S-tab:alternatives} reports paired changes relative to a control that omits the tested modification. Each row compares the same ten inputs within one session. The three stopping-rule controls retain direct integration, the integration control uses DCT solves, the residue-balanced control uses midpoint binary cuts, and the early-exit controls use a full fused residue table. All other choices are held fixed within each pair.

\begin{table}[!htbp]
\centering\small
\caption{Paired effects of implementation variants on ten inputs at minimum side $s$ pixels. Time ratios compare each variant with the control specified below, using per-input median timings; the table reports the median ratio across inputs. Values above one are slower. $\Delta\SP$ and $\Delta$RMSE are mean paired changes, variant minus control, in percentage points (pp) and radians. $\SP$ is the fraction of pixels with absolute reference error below $\pi$ after mean-offset removal. Positive $\Delta\SP$ and negative $\Delta$RMSE indicate improvement. $N$ is the processed pixel count. Rows from different timing sessions do not establish a ranking between variants.}\label{S-tab:alternatives}
\begin{tabular}{lrrrrrr}
\toprule
Variant & \multicolumn{2}{c}{Time ratio} & \multicolumn{2}{c}{$\Delta\SP$ (pp)} & \multicolumn{2}{c}{$\Delta$RMSE (rad)} \\
& $s=8$ & $s=16$ & $s=8$ & $s=16$ & $s=8$ & $s=16$ \\
\midrule
Density $0.001$ & 1.027 & 1.016 & -5.93 & -0.78 & +0.240 & +0.090 \\
Density $0.01$ & 1.061 & 1.124 & -1.52 & -1.30 & +0.556 & +0.016 \\
Low-charge cutoff & 0.991 & 1.023 & -1.44 & +0.46 & +0.873 & +0.106 \\
Verified integration & 1.039 & 1.012 & -0.04 & 0.00 & +0.002 & 0.000 \\
Residue-balanced kd-tree & 1.525 & 1.847 & -0.70 & -3.07 & +0.420 & -0.190 \\
Early exit, cap $N$ & 1.038 & 1.033 & 0.00 & 0.00 & 0.000 & 0.000 \\
Early exit, cap $2N$ & 1.041 & 1.029 & 0.00 & 0.00 & 0.000 & 0.000 \\
Early exit, cap $4N$ & 1.038 & 1.028 & 0.00 & 0.00 & 0.000 & 0.000 \\
\bottomrule
\end{tabular}

\par\smallskip
\begin{minipage}{\textwidth}\footnotesize
\textit{Controls.} Density and low-charge rules: residue-count quadtree with direct integration. Verified integration: quadtree with DCT solves. Residue-balanced cuts: midpoint kd-tree. Early exit: quadtree with a full fused residue table.

\textit{Timing.} Density, low-charge, and binary-cut tests use five repeats; integration uses seven; early exit uses nine. The first five rows exclude temporary-buffer destruction; the last three include it. Rounded zero metric changes do not imply identical output arrays; exact identity was verified separately for early exit.
\end{minipage}
\end{table}

\paragraph{Relaxed residue stopping.}
A density rule stops a tile when residue count divided by the number of interior cells is at most 0.001 or 0.01. A low-charge rule stops when the count is below $0.25(h+w)$ on a tile whose largest side is at most $4s$. These rules can retain inconsistent tiles, unlike zero-residue stopping. They were compared using the same integration-enabled local solver. Neither consistently improved both measured time and reference agreement at $s=8$ and $s=16$, so neither was retained for the main comparison.

\paragraph{Verified direct integration.}
A candidate solution is formed by integrating observed differences, then checked against every internal edge. The DCT can be skipped only when those differences agree with the integrated field. The check itself costs work, and the change in arithmetic and quantization can affect the output. The grid is given the same integration option. The tested version did not consistently reduce measured reconstruction time, so the selected pipeline keeps the DCT solve on every tile.

\paragraph{Early-exit residue search.}
For unrestricted residue refinement, only existence of a nonzero residue is needed. A scan can stop as soon as one is found, but repeated parent/child scans revisit cells. We tested limits of $N$, $2N$, and $4N$ inspected cells before reverting to a summed-area table, where $N$ is the processed pixel count. Tested partitions and outputs were preserved, but the extra scans were slower overall. The main implementation computes the fused residue table once.

\paragraph{Residue-balanced binary cuts.}
A summed-area table supports binary searches for a cut near half the residue count along an eligible axis. The cut is clamped to preserve the minimum child size. Compared with the midpoint control used in this test, the search adds work and may produce less favorable transform shapes. The exploratory results did not establish a consistent benefit, so the main comparison uses the two geometric rules.

\section{Implementation and numerical checks}\label{S-sec:validation}
The complete reconstruction experiments use a frozen BRiDCT~\cite{moevus2026bridct} version dated 20 September 2026 for the grid and both tree geometries. Every method receives the same encoded and padded arrays. BRiDCT uses single precision; phase preparation and boundary statistics use double precision. FFTW 3.3.11 uses \texttt{FFTW\_ESTIMATE} plans for unsupported shapes, distinct from the measured plans in the transform microbenchmark. Forward and inverse scaling is matched, and the constant Poisson coefficient is zero. Transform implementations are frozen for each timing session.

To separate intended algorithm changes from numerical side effects, we check partitions and reconstructed outputs as well as timing. The main timing comparison on 198 entries contains $198\times5\times2\times7=13{,}860$ timed calls: a grid, the quadtree before the optimizations defined in the main report (Sec.~\ref{M-sec:optimized}), and three optimized adaptive configurations at two minimum sizes. Warm-ups and correctness checks are additional. For each image, numerical checks compare the residue maps, partitions, and outputs before and after the optimizations intended to preserve them. These construction and assembly checks cover fused residue preprocessing, interface collection, offset propagation, and the early-exit search. Across this set, 2772 comparisons verify identical partitions and outputs for the tested combinations. These checks do not imply identity after replacing a transform fallback: final rounding differs from the reference quadtree on 15 entries at $s=8$ and eight at $s=16$, although success fractions remain unchanged and the largest absolute RMSE change is below $5.1\times10^{-4}$ radians. Reconstruction metrics are therefore checked explicitly rather than inferred from bitwise equality.

The 256-level encoding requires a consistent convention for half-cycle edge differences. A focused diagnostic on 256 constructed cells found no disagreement between the retained residue zero/nonzero test and the integer local-gradient circulation for those patterns. This is a finite endpoint check, not a proof for arbitrary observations.

\section{First-call latency}\label{S-sec:cold}
The main report measures repeated reconstruction with initialized transform plans and reusable buffers. Table~\ref{S-tab:cold} measures the first reconstruction in a fresh process, which additionally includes initialization and memory/cache effects within the reconstruction call. Process startup, input loading, and encoding remain outside the timer.

This first-call study uses $10\times3\times2\times3=180$ fresh processes, crossing inputs, methods, minimum sizes, and process repetitions. Each performs one first call and seven warm calls; all 1260 warm outputs equal their process's first-call output. First-call times are summarized across three processes per image/configuration. Warm times use the median within each process, then across processes. Paired grid ratios are formed only after these within-case summaries. No timing sessions are pooled.

\begin{table}[!htbp]
\centering\small
\caption{First-call overhead and adaptive/grid comparison on the ten exploratory inputs. $s$ is the minimum adaptive side, or nominal grid side, in pixels. Each entry/configuration uses three fresh processes, each with one first call and seven warm calls. First-call time is the median over the three processes; warm time is the median of their seven-call medians. Columns report medians of the resulting per-entry ratios. First/warm compares the same method before and after initialization; first/grid compares its first call with the grid's first call at the same $s$. Ratios above one mean longer reconstruction. Process startup, input loading, and encoding are excluded.}\label{S-tab:cold}
\begin{tabular}{llrr}
\toprule
$s$ (px) & Partition & First/warm & First/grid \\
\midrule
8 & Grid & 2.245 & 1.000 \\
8 & Quadtree & 2.151 & 1.219 \\
8 & KD dyadic & 2.244 & 1.244 \\
16 & Grid & 2.938 & 1.000 \\
16 & Quadtree & 2.391 & 1.378 \\
16 & KD dyadic & 2.459 & 1.373 \\
\bottomrule
\end{tabular}

\end{table}

\section*{Availability and interpretation}
Public sources identify the underlying collections but do not reproduce every derived array. Some photograph metadata lack an exact public filename, and some generated fields lack a complete retained recipe. The benchmark arrays, implementations, and outputs are not currently publicly available. These restrictions limit independent reproduction of the numerical results. The descriptions above specify the evaluated operations, inputs, and controls, but a public release is still needed for independent replication.

\end{document}